\documentclass[5p,twocolumn,times]{elsarticle}

\usepackage{amssymb}
\usepackage{lineno}
\usepackage{amsmath}
\usepackage[dvipsnames]{xcolor}
\usepackage{hyperref}

\definecolor{darkpastelgreen}{rgb}{0.01, 0.75, 0.24}

\newcommand{\nn}{\nonumber}

\newcommand{\derc}[3]{\displaystyle \left( \frac{\partial #1}{\partial #2} \right)
\raisebox{-1em}{\ensuremath{#3}}}
\def\eps{\varepsilon}
\def\fmc{\rm ~fm^{-3}} 
\def\fm{\rm ~fm}
\def\fm2{\rm ~fm^2} 
\def\MeV{\rm ~MeV}    
\def\mL{{\cal L}}
\def\Cs{C_\sigma^2}
\def\Co{C_\omega^2}
\def\Cd{C_\delta^2}
\def\Cr{C_\rho^2}
\def\Lsd{\Lambda_{\sigma\delta}}
\def\Lod{\Lambda_{\omega\delta}}
\def\pa{\partial}

\newcommand{\correction}[1]{{#1}}

\graphicspath{{./figs/}}

\journal{Physics Letters B}

\begin{document}

\begin{frontmatter}

\title{Examining the possibility of a thick crust in the compact stellar
    object HESS J1731-347}

\author{Sebastian Kubis}
\ead{skubis@pk.edu.pl}

\author{Włodzimierz Wójcik}

\affiliation{organization={Cracow University of Technology}, 
            addressline={Department of Physics, Podchorażych  1},
            city={Kraków},
            postcode={30-084}, 
            country={Poland}}

\begin{abstract}
A relativistic mean-field model with a crossing term of isovector-scalar and isoscalar
mesons,
the Cracow crossing terms
(CCT) model, has previously been shown to be capable of explaining the light and compact
object
HESS J1731-347 \cite{Kubis:2023gxa}.
Here, the model is supplemented with a correct treatment of the phase transition. 
Applying the thermodynamically consistent Gibbs conditions shows that a mixed-phase system
occurs
in the core of the neutron star over a wide range of densities, extending up to the
neutron star crust,
leading to an intriguing stellar structure with a thick and massive crust.
\end{abstract}


\end{frontmatter}

\section{Introduction\label{sec:intro}}

The supernova remnant HESS J1731-347~\cite{Doroshenko:2022} appears to contain a neutron
star (NS) with an intriguingly
compact size (radius around 10~km) and low mass (around $0.8~M_\odot$). 
It is expected that a low-mass star would have a larger radius because of its weaker
gravity.
Such a compact size for a low-mass NS may only be explained by a rather specific equation
of state (EOS).
This EOS must describe matter that is soft at low densities (around saturation density,
$n_0$) and stiffens rapidly as the density increases to
support 
the existence of a star with mass above $2~M_\odot$, in line with the well-constrained
observation of the most massive NS (PSR J0740+6620~\cite{Fonseca:2021wxt}) at $2.08 \pm
0.07~M_\odot$. This leads to a characteristic Z shape in the mass--radius relation, with
lower-mass stars being more compact
.

Many models of compact stars have been proposed to explain  HESS J1731-347. Most involve
exotic components that are responsible for softening the EOS, like kaon
condensation~\cite{Veselsky:2024eae}
or strange quark matter~\cite{Laskos-Patkos:2023tlr}, which lead to a hybrid star. The EOS
softening may be a result of a phase
transition, regardless of the type of matter in which it occurs. A general analysis of
that kind was presented in~\cite{Laskos-Patkos:2024otk}. Exotic components may explain the
compact size of HESS J1731-347, but these approaches do not generally align well with
other mass--radius measurements or estimates based on
gravitational wave detection. 

Some authors have instead attempted to revise the standard nuclear models, without
introducing exotic components. A promising idea was to include meson crossing terms in the
relativistic mean-field theory.
The scalar-meson $\sigma$--$\delta$ crossing term, introduced in~\cite{Zabari:2019ukk},
leads to a soft EOS
at low densities. The idea of a scalar-meson crossing term together with vector-mesons crossing terms 
was later considered in~\cite{Miyatsu:2022wuy,Xia:2024nwa,Li:2022okx}. Many meson crossing terms were
thoroughly explored in~\cite{Miyatsu:2024ioc}.
Thanks to these works, one may conclude that the inclusion of 
vector-vector meson crossing terms
eliminates the desired softening of matter at low densities. 
For this reason, in the present work, we neglect vector-vector terms, retaining 
only scalar-vector and scalar-scalar crossing terms.
Another valuable class of approaches is the density-dependent relativistic mean-field
(DD-RMF) models.
In~\cite{Xia:2024nwa}, with a 
Bayesian analysis, it was shown that DD-RMF models without meson crossing terms cannot
agree with HESS J1731-347 and other
observational constraints.  
Additionally, nonstandard explanations for HESS J1731-347 have been proposed, like a
neutron star with a dark matter component~\cite{Hong:2024sey} and extended gravity, in
which the standard Tolman--Oppenheimer--Volkoff equations take a different
form~\cite{Lope-Oter:2024egz}.

In~\cite{Kubis:2023gxa}, a nuclear model was proposed, which has an EOS with the required
properties and can accommodate low-mass and compact objects like HESS J1731-347 alongside
other mass--radius constraints from observations. In this model, the dense matter
undergoes a phase transition at a density slightly
above the saturation point. To obtain a unique pressure--density relation, the commonly
used Maxwell construction was employed. It is
well known 
that the Maxwell construction is not quite correct for nuclear matter with leptons. It was
discussed thoroughly in the chapter 9.2 of
~\cite{Glendenning:1997wn} that for systems with two conserved quantum numbers, charge and
baryon number, the Maxwell construction cannot ensure chemical equilibrium for the two
separated phases. The pressure and chemical
potentials for all particle species $i$ present in both phases, denoted by $I$ and $II$,
must be the same:
\begin{equation}
 P^{(I)} = P^{(II)} ~,~~ \mu_i^{(I)} = \mu_i^{(II)}.
 \label{Gibbs-cond}
\end{equation}
The Maxwell construction, also applied in~\cite{Kubis:2023gxa}, is based on the assumption
that neutron--proton--lepton ($npl$) matter is
locally neutral. Under this constraint, we cannot ensure that all the Gibbs conditions are
satisfied. The {\it equal-area} rule,
in the case of $npl$ matter, breaks the equality of chemical potentials for all charged
particles, that is, $\mu_p, \mu_e$, and
$\mu_\mu$.

In this work, we use the so-called Gibbs construction, in which the proper conditions for
the two-phase system are fulfilled.
It is found that the matter undergoes a phase separation, with 
one phase free of protons. This coexistence of an $npl$ phase with a neutron--lepton
($nl$) phase spans a broad range
of densities, from a moderate NS core density down to the deep inner crust region. This
means that the NS crust is thicker and
makes a substantially greater contribution to the total stellar mass and moment of inertia
than in typical approaches. Our aim is to present this
intriguing NS structure and investigate possible observational consequences.

\section{The relativistic mean-field model}

The Cracow crossing terms (CCT) model of nuclear interaction proposed
in~\cite{Kubis:2023gxa} is based on a relativistic Lagrangian
with a standard kinetic term and interaction terms:
\begin{equation}
\mL = \mL_\text{kin} + \mL_{N\phi} + \mL_\text{cross} ~,~~ \phi=\sigma,\omega,\rho,\delta
.
\label{Ltot}
\end{equation}
It uses four types of meson fields, $\phi = \sigma, \omega, \rho, \delta$, 
which are linearly coupled to all nucleonic currents admissible by the isospin symmetry:
\begin{equation}
\mathcal{L}_{N\phi} = g_\sigma\sigma\bar{\psi}\psi
-g_\omega\omega_\mu\bar{\psi}\gamma^\mu\psi -
\frac{1}{2}g_{\rho}\vec{\rho}_{\mu}\bar{\psi}\gamma^\mu\vec{\tau}\psi+g_\delta\vec{\delta}\bar{\psi}\vec{\tau}\psi
-U(\sigma),
\label{L-int}
\end{equation}
where 
\begin{equation}
U\left(\sigma\right)=\frac{1}{3}b m {(g_\sigma\sigma)}^3+\frac{1}{4}c {(g_\sigma\sigma)}^4\end{equation}
is the self-interaction of the $\sigma$ meson. 
The key ingredients of the model are the meson--meson interaction terms (crossing terms),
in which the
scalar-isovector field $\vec{\delta}$
is  coupled to the two isoscalar fields, $\sigma$ and $\omega$:
 \begin{equation}
\mL_\text{cross} =
 \frac{1}{2} g_{\sigma\delta}\, \sigma^2\vec{\delta}^{\,2} +  \frac{1}{2}
g_{\omega\delta}\,\omega_\mu\omega^\mu \vec{\delta}^{\,2} .
\label{L-cross-terms}
\end{equation}
In the spirit of the relativistic mean-field approach, the meson fields contribute to the
total energy through their average values, which can then
be replaced by the effective masses of nucleons:
\begin{align}
m_p = m-g_\sigma\bar{\sigma}-g_\delta{\bar{\delta}}^{(3)} ,\label{mp} \\
m_n = m-g_\sigma\bar{\sigma}+g_\delta{\bar{\delta}}^{(3)} , \label{mn}
\end{align}
where $m$ is the bare nucleon mass.
 
The energy density for nucleonic matter can be expressed in terms of the nucleon scalar
$n_i^s$ and vector $n_i$ densities and effective
masses, which are functions of the densities $m_i(n_p,n_n)$, for $i=p,n$:
\begin{align}
\eps_\text{nuc} = &  \sum_{i=p,n}\frac{1}{4}(3 E_{F,i} n_i + m_i n_i^s) +
   \frac{1}{2 \Cs}(m-\bar{m})^2 \nonumber \\
 & + \frac{\Co}{2} \frac{n^2}{1 + \Co \Lod (\Delta m/2)^2} +
  \frac{\Cr}{8} (2x-1)^2 n^2 \nonumber \\
 & + \frac{\Delta m^2}{8 \Cd} +  \frac{1}{8}\Lsd (m-\bar{m})^2\Delta m^2 + U(m-\bar{m}),
 \label{enuc}
\end{align}
where $\Delta m = m_n -m_p$ , $\bar{m}= (m_n+m_p)/2 $ {and $x=n_p/n$}. Similarly, the
pressure is
\begin{align}
P_\text{nuc} = & \sum_{i=n,p}\frac{1}{4}\left(n_i E_{F,i} - m_i n_i^s\right)
-\frac{(m-\bar{m})^2}{2 \Cs} \nn \\
& + \frac{\Co n^2}{2 \left(\frac{1}{4} \Co \Delta m^2\Lod + 1\right)} + \frac{1}{8} \Cr
n^2 (1-2 x)^2 \nn \\
& - \frac{\Delta m^2}{8 \Cd} + \frac{1}{8} \Delta m^2 \Lsd (m-\bar{m})^2 - U(m-\bar{m})
  \label{Pnuc}
\end{align}
with a typical negative contribution from scalar mesons.
In the expressions for the energy density and pressure, it is more convenient to introduce
coupling constants different from those appearing in the
Lagrangian: $C_i = g_i/m_i$, $\Lsd = g_{\sigma\delta}/g_\sigma^2 g_\delta^2$,
and $\Lod = g_{\omega\delta}/g_\omega^2 g_\delta^2$. Together with $b$ and $c$ in the
$U(\sigma)$ potential, they are independent parameters which uniquely define the
model. Their values come from fitting to the basic properties of matter at the saturation
point density $n_0 = 0.16 \fmc$,
like the binding energy per nucleon $E_0 = - 16.0 $~MeV, the compressibility of symmetric
matter $K_0 = 230 \MeV$,
the symmetry energy $E_{sym} = 30 \MeV$. A discussion of the model parametrization is
presented in \cite{Kubis:2023gxa}.
In that work it was shown, how the NS radius depends on the symmetry energy slope $L$. The
values between 40 and 80~{\rm MeV} were
examined and only low values of $L$, around $40 \MeV$, ensure the radius 
sufficiently small for \correction{a low-mass star}, as in the HESS J1731-347 case.
Hence, in this work, we leave the only one value of $L=40 \MeV$ but keep three different
$\Cs$.
Changes of the $\Cs$ values have visible effect on the NS radius and they 
are also relevant for the maximum mass of the NS. Another reason of keeping $L=40 \MeV$ is that the
phase separation region diminishes quickly
with increasing $L$ and vanish completely around $L\approx 60 \MeV$.
Table~\ref{tab-const} presents parameters for three models with $L=40 \MeV$ and different
stiffness at higher densities.
A detailed discussion of model parametrization and its connection to experimental data as
well as the the constraints coming from NS
observations is presented in our previous works~\cite{Zabari:2019ukk}
and~\cite{Kubis:2023gxa}.
\begin{table}[t!]
\caption{Coupling constants for CCT models with $L=40 \MeV$ and different values of
$\Cs$.}
\small
\begin{tabular}{ccccccc}
\hline\hline
$\Cs$       &  $\Co$ & $b$ & $c$				   & $\Cr$  & $\Cd$ \\ 
(fm$^2$)    & (fm$^2$) & (-)	& (-) 					   & (fm$^2$) & (fm$^2$) \\ \hline
 12 	    & 6.9769 ~  & 0.004733 & -0.0052878 & 15.2938 &  2.63799 \\
 13  	    & 7.9531 ~  & 0.003695 & -0.0045224 & 13.9852 & 2.33674 \\ 
 14  	    & 8.9055 ~  & 0.003005 & -0.0039370 & 12.9888 & 2.11461 \\ \hline\hline
\end{tabular}
\label{tab-const}
\end{table}

\section{Phase separation}

In addition to nucleons, the NS matter contains the leptons $e$ and $\mu$, which are
produced in the following beta processes:
\begin{equation}
n \rightarrow p + l +\bar{\nu}_l ~,~~ p + l \rightarrow n + \nu_l.
\label{betaeq}  
\end{equation}
The lepton contributions to the total energy and pressure are given by standard
expressions that depend only on the lepton chemical potentials $\eps_{i}(\mu_i)$ and
$P_i(\mu_i)$, for $i=e,\mu$.
As discussed in~\cite{Kubis:2020ysv}, for low values of $L$, the $npl$ matter becomes
unstable with respect to charge
fluctuations and undergoes a phase transition. The quantity that indicates the instability
of the matter is its incompressibility under a
constant chemical potential:
\begin{equation}
K_\mu = \derc{P}{n}{\mu} > 0 .
\label{Kmstab} 
\end{equation}
In the region of densities where this condition is violated 
{(the so-called {\it spinodal} region)}, the matter cannot stay as a one-phase system; it
must split into two coexisting phases with different charge and baryon densities. The
coexistence conditions~(\ref{Gibbs-cond}) require the
equality of chemical potentials (chemical equilibrium) and pressures (mechanical
equilibrium) between those two phases. Usually, the NS
matter remains in beta equilibrium at all densities. The temperature of NS matter can be
on the order of millions of
kelvins and is negligible in comparison to the chemical potentials, $k_B T \ll \mu_i$, but
this means that the energy levels around the Fermi
energies (the Fermi energy is here equal to the chemical potential) are not completely
occupied.
So, the particles may exchange their energy via the beta reactions~(\ref{betaeq}), which
leads to the following relation between chemical potentials for the degenerate fermions:
\begin{equation}
  \mu_l = \mu_n -\mu_p.
\end{equation} 
The chemical potential $\mu_\nu$ is absent here, as the weakly interacting neutrinos
escape from the system, carrying away
thermal energy on the order of $k_B T$. The symmetry energy 
{$E_s = \frac{1}{8}\frac{\pa(\eps/n)}{\pa^2 x}|_{x=1/2} $} uniquely determines the proton
fraction $x$ and the lepton
abundance through the relation {which is valid within the so-called parabolic
approximation \cite{Lattimer:2006xb}.}
\begin{equation}
 \mu_n -\mu_p = 4(1-2x)E_{s} .
\end{equation}
Thus, when $E_s(n) \le 0$, lepton production is blocked and the nucleonic matter contains
only
neutrons. However, the neutron phase may also appear for $E_s(n) >0$ as a consequence of
the Gibbs conditions
for a two-phase system. When the symmetry energy is low but still positive, it may occur
that one of the phases (the one with lower
density) cannot contain protons, but only neutrons and leptons, because the difference
between neutron and lepton energies
is too small to produce protons:
\begin{equation}
  \mu_n^I -\mu_l^I < \mu_p^I.
\end{equation}
Such a behaviour was discussed in~\cite{Kubis:2007zz}. 
This is a mechanism analogous to the neutron drip phenomenon occurring in the inner crust
of the NS
\cite{Douchin:2000kad}.
In our model, it appears that the Gibbs conditions are satisfied for neutrons and
electrons but not for protons:
\begin{equation}
\mu_n^I = \mu_n^{II}, ~~ \mu_l^I = \mu_l^{II} , ~~ \mu_p^I > \mu_p^{II}. 
\end{equation}
Protons are confined to the $II$ phase as they have lower energy there.

\begin{figure}[t]
\includegraphics[width=1\columnwidth]{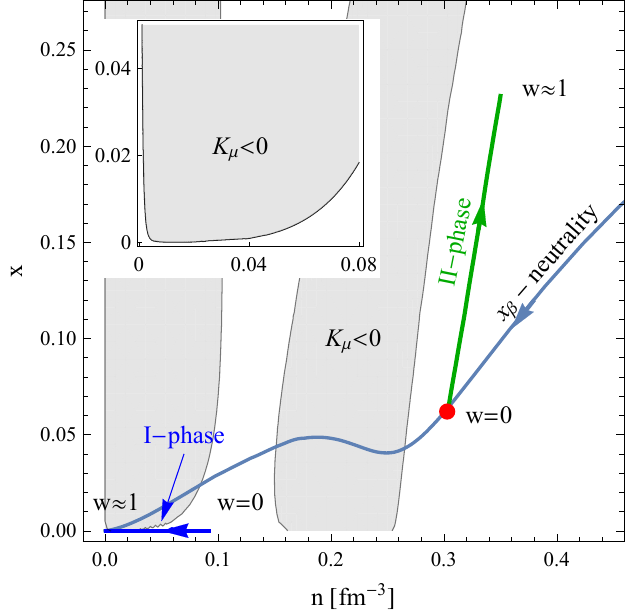}
	\caption{\label{phase-diagram} 
Phase coexistence diagram for the model with $\Cs =12$ and $L=40$. The spinodal (gray)
regions correspond to $K_\mu<0$.
Blue and green lines depict $I$ and $II$ phases, respectively, with the limiting values of
$w$. The red point indicates the place
	where the homogeneous phase vanishes in favor of the two-phase system. 
{The inset demonstrates that at low densities the spinodal line does not reach $x = 0$,
confirming the stability of phase
	$I$.}}
\end{figure}

Summarizing, the system consists of two phases: a low-density phase ($I$)
with neutrons and leptons and a high-density phase ($II$) 
containing neutrons, protons, and leptons. 
The $I$ phase is negatively charged, whereas the $II$ phase is positively
charged. The phases must fill the space in the following ratio to ensure the global
neutrality of the dense matter:
\begin{equation}
- w (n_e^I+n_\mu^I) + (1-w)(n_p^{II} -n_e^{II} -n_\mu^{II})  = 0~ ,
\end{equation}
where $w=V^I/V$ is the low-density volume fraction and $V=V^I+V^{II}$ is the total volume
of the system.
Figure~\ref{phase-diagram} shows the phase diagram of the nucleon--lepton matter. The
light-blue line
represents the proton fraction $x_\beta$ for a homogeneous, one-phase system under beta
equilibrium that is charge neutral.
 This neutrality line marks the border between positively charged matter 
(all $(n,x)$ points above) and negatively charged matter (below). 
The neutrality line passes through unstable regions $K_\mu <0$, so it cannot 
represent a physical system. Going from high densities, where neutral matter is stable, at
some density around $0.3 \fmc$,
the system splits into two phases, which are depicted by blue and green lines. The
proton--neutron phase $(II)$ fades in favour of the
neutron phase $(I)$, as $w \rightarrow 1$; however, the I-phase never fills the entire
space, even for low densities.
Therefore, at low densities, $w$ is only approximately equal to \correction{1}. 

The average baryon number density of the two-phase system is given by
\begin{equation}
  \bar{n} = w\, n^I + (1-w) n^{II}
\end{equation}
and so, the  energy of the two-phase system is the weighted  sum of both phases:
\begin{equation}
  \bar{\eps} =  w\, \eps^I + (1-w) \eps^{II}.
\end{equation}

\begin{figure}[t]
\includegraphics[width=.99\columnwidth]{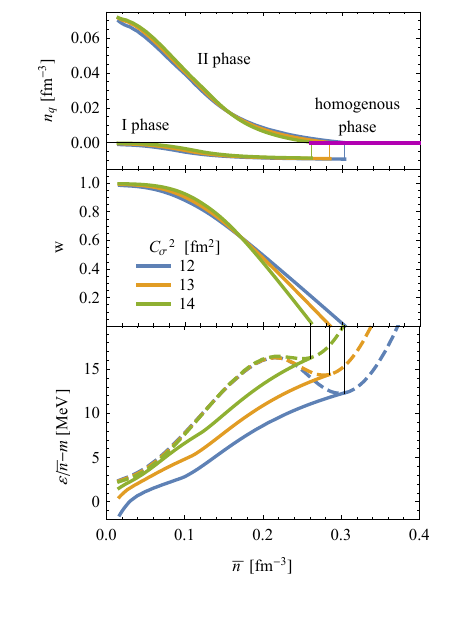}
	\caption{\label{endiffw-plot}
Lower panel: Energy per particle for homogeneous matter (dashed lines) and for the
two-phase system (solid lines).
Upper panels: Volume fraction $w$ and charge density $n_q$ of both phases for the three
different nuclear models.}
\end{figure}

Figure~\ref{endiffw-plot} shows the energy of the two-phase system and that 
of homogeneous, $\beta$-equilibrated matter for three different models. It confirms that
the two-phase
system is the preferred state of matter at all densities below $n\approx 0.3 \fmc$.  
A similar two-phase system appearing at densities slightly above $n_0$ was analyzed
in~\cite{Kubis:2020ysv}, but there, the two-phase
system disappeared close to $n_0$ and $npl$ matter was again homogeneous. Here, the
two-phase matter persists down to
all densities below $n_0$. The residual proton--neutron phase may be interpreted as 
quasi-nuclei sparsely placed in the region dominated by neutron fluid; such a system
corresponds to the inner crust structure.
So, in our model, the crust--core transition moves up to a density around $0.3 \fmc$. This
is much higher than the typical
crust--core transition point, which, depending on the nuclear model, fluctuates in the
range between 0.03 and 0.08 $\fmc$.
In this work, we employ the bulk approximation - finite-size effects such as surface and Coulomb
energies are not included. These effects play an essential role in determining the size of 
non-uniform structures (droplets, rods, slabs, etc.) and reduce the density range
over which a mixed phase can exist. 
In this way, the high-density limit of the two-phase system $n_{2ph}$ may be treated only as an upper
 bound for  the true crust-core transition $n_{cc}$.
However, the difference between $n_{cc}$ and $n_{2ph}$ appears to be small. 
In particular, at densities of about 0.3 $\fmc$, the electron chemical potential is much larger 
($\sim 100 \MeV$) than the Coulomb or surface energies (a few MeV) arising from the density discontinuity
 at the phase boundary. Consequently, throughout this work we treat $n_{2ph}$ 
 as the effective crust–core transition density. Table~\ref{nc-tab}
 lists the corresponding critical densities for the formation of the two-phase system. 
 The transition density decreases slightly with increasing strength of the $\Cs$ coupling.
\begin{table}[h]
\caption{Values of the critical density for two-phase system formation for three different
models.}
\center
 \begin{tabular}{cccc}
 \hline\hline
  $\Cs$ (fm$^2$) & 12 &  13 & 14   \\ \hline
  $n_{\rm 2ph}$ (fm$^{-3}$) & 0.3038 & 0.2849 & 0.2610 \\
  $n_\text{join}$ (fm$^{-3}$)  & 0.0025 & 0.0656 &  0.0656 \\  \hline\hline
 \end{tabular} 
 \label{nc-tab}
\end{table}

\correction{
The CCT model is not adequate for the very low-density regime 
(e.g., the description of the outer crust is based on experimental masses of neutron-rich nuclei 
\cite{Haensel:1993zw}).
In this regime, we adopt the commonly used EOS based on the SLy nuclear model \cite{Douchin:2001sv} 
to describe $P(\rho)$ in the outermost layers of the stellar crust.
The two EOSs intersect at a certain density, defined by $P_{SLy}(n_\text{join}) = P_{CCT}(n_\text{join})$.
Thus, below $n_\text{join}$ we use $P_{SLy}$, while above $n_\text{join}$ we use $P_{CCT}$.
In this manner, the two EOSs are "stitched" at $n_\text{join}$, which ensures mechanical 
equilibrium between the two regions. The values of these joining densities, $n_\text{join}$, 
are given in Table~\ref{nc-tab}. It is worth noting that $n_\text{join}$ is always located within 
the inner-crust region of the SLy EOS, which is structurally consistent with the two-phase
system of CCT, consisting of proton clusters immersed in a neutron-lepton liquid.
}

\begin{figure}[t]
\includegraphics[width=\columnwidth]{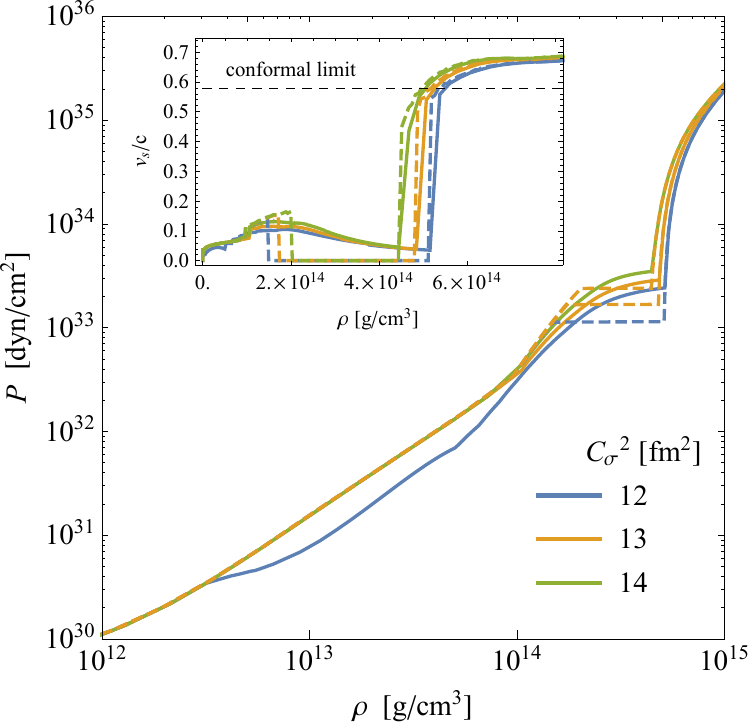}
\caption{\label{Fig:EOS}
Plot of the EOS for three CCT models. The inset shows the speed of sound in the
two‑phase region. Solid and dashed lines correspond to EOSs obtained using the
Gibbs and Maxwell constructions, respectively.}
\end{figure}

The EOSs for all three  $\Cs$ values are shown in Fig.~\ref{Fig:EOS}.
For comparison, the EOSs based on the Maxwell construction are shown with dashed lines.
 One may notice that the Gibbs construction substantially alters the 
pressure--density relation. Especially for the model with $\Cs = 12$, the EOS is much
softer at low densities.
In Fig.~\ref{Fig:EOS}, a plot with the speed of sound, defined as
$v_s=\sqrt{\derc{P}{\rho}{q}}$, is also shown. The dashed lines extend into
the unphysical region where $v_s$ vanishes;
a vanishing speed of sound leads to a discontinuity in the density profile of an NS.
The solid lines, corresponding to the Gibbs construction, indicate small but finite values
of $v_s$ at lower
densities, which rapidly increase and even exceed the limit $v_s < c/\sqrt{3}$ from the
EOS for
free ultra-relativistic fermions~\cite{Bedaque:2014sqa}, also called {\em the conformal limit}.
 Various limits for $v_s$ were discussed recently
in~\cite{Laskos-Patkos:2024otk}; it was shown that the $c/\sqrt{3}$ limit must be broken
if the nuclear EOS
 is to explain HESS J1731-347 observations and the fact that $M_\text{max}$ for an NS is
above $2~M_{\odot}$. The values of $v_s$ obtained here are in agreement with the more
general condition:
\begin{equation} 
v_s < \frac{\eps - 1/3 P}{\eps + P},
\end{equation}
which is valid for any $P(\eps)$ and comes from relativistic kinetic
theory~\cite{Olson:2000vx}.

The key finding of this section is that a nuclear model with a correct chemical
equilibrium condition for the two-phase system leads to a thick NS crust.
A similar scenario, in which a $nl$ phase coexists with an $npl$ phase, was considered
in~\cite{Kubis:2020ysv}, but there,
 the two-phase system occurs at densities 
approximately between $0.1 \fmc$ and $0.3 \fmc$, 
corresponding to the outer core region. 

{Such non‑typical behaviour arises mainly from the non‑monotonicity of the
symmetry energy, as already discussed in \cite{Kubis:2020ysv}. When the symmetry
energy decreases with density and approaches zero, it leads to a violation of
the stability conditions for a homogeneous system \cite{Kubis:2006kb}. Moreover,
if the symmetry energy becomes large and negative, it results in the formation
of a local minimum in the energy of asymmetric matter (see Fig. 2 in
\cite{Kubis:2020ysv}). This minimum corresponds to a new state of dense matter,
already proposed in \cite{Lee:1974ma} and later referred to as the {\it density
isomer}. Possible signals of the density isomer in symmetric matter were
considered in \cite{Stoecker:1979jn, Nix:1981ta}. 
In our model, the density isomer appears in highly asymmetric matter, and it
would be interesting to estimate its possible effects in future heavy‑ion collision
experiments involving highly asymmetric nuclei, such as FAIR or NICA. 
 
At this point we would like to refer to the non‑monotonicity of $E_{\text{sym}}$ 
in the context of
recent experiments measuring the neutron skin thickness, namely PREX‑II
\cite{PREX:2021umo} and CREX \cite{CREX:2022kgg}. The neutron skin thickness
extracted from the PREX-II experiment is significantly larger than that obtained in
CREX, and consequently implies much larger values of the slope parameter $L$ than
the latter.
These results have triggered an intense discussion concerning the value of $L$
and other symmetry energy parameters
\cite{Sammarruca:2023mxp,Zhang:2022bni,Reed:2023cap,Burgio:2024vjc}. The model
presented here favours relatively low values of $L$. There are also suggestions
\cite{Kumar:2024bud,Miyatsu:2023lki} that the inclusion of $\sigma\!-\!\delta$
coupling in RMF models may help reconcile the PREX–CREX discrepancy.

Another important discussion is presented in \cite{Lattimer:2023rpe}, where it is shown
that the neutron skin thickness correlates not only with $L$, but also with
$K_{\mathrm{sym}}$, the curvature of $E_{\mathrm{sym}}(n)$ at $n_0$.
That work indicates that reconciling the PREX‑II result with low
values of $L$ requires a negative $K_{\mathrm{sym}}$. A similar conclusion was
reached in \cite{Guan:2025yia}, where the discrepancy between PREX‑II and the
GW170817 observation was resolved by assuming negative values of
$K_{\mathrm{sym}}$.

At present, no firm conclusion on $L$ and $K_{\mathrm{sym}}$ has been
established; however, low $L$ and negative $K_{\mathrm{sym}}$ naturally favour
a non‑monotonic behaviour of $E_{\mathrm{sym}}(n)$ appearing in CCT models. Such
features - like low $L$, negative $K_{\mathrm{sym}}$, and even non‑monotonic
symmetry energy - also emerge in chiral approaches
\cite{Drischler:2020yad,Tews:2024owl}. Notably, the non‑monotonic behaviour of the
symmetry energy is a key ingredient leading to a two‑phase structure that extends
to significantly lower densities, reaching the region of the inner crust.
}
It seems natural to conclude that in this case, the NS crust extends to much deeper
regions of the star.
In the next section, we consider the implications for NS observables.

\section{Neutron star properties}
 
\begin{figure}[t]
\includegraphics[width=1\columnwidth]{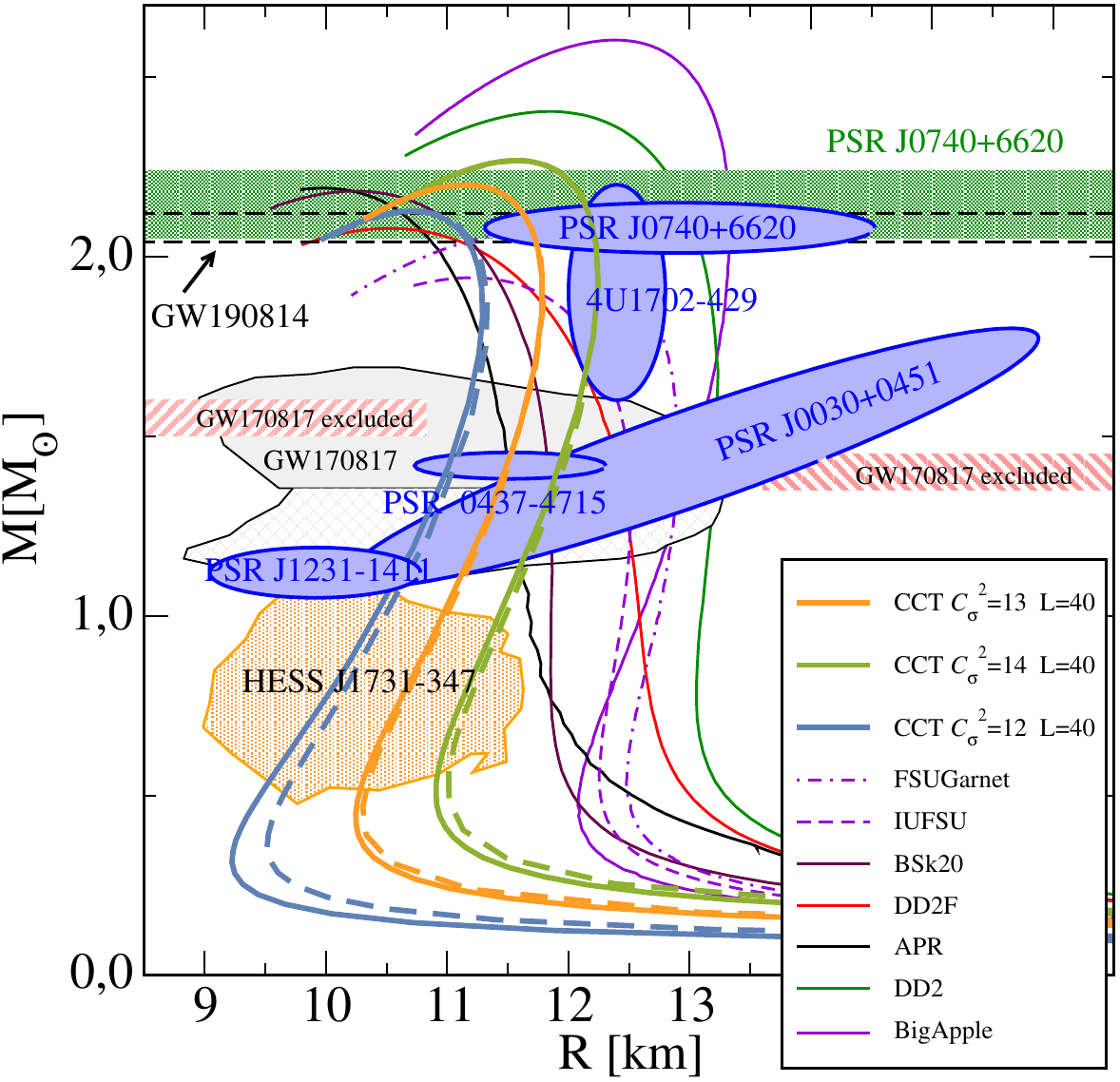}
	\caption{\label{Fig:MR}
Mass–radius relations for CCT models  with $L=40$ MeV. Thick solid and dashed
lines correspond to EOSs constructed using the Gibbs and Maxwell constructions,
respectively. Selected nuclear models are shown for comparison (thin lines). Shaded
regions indicate observational constraints from pulsars and gravitational‑wave
detections.}
\end{figure}

In this section, the most relevant NS properties are discussed with respect to the
inclusion of Gibbs conditions.
In Fig.~\ref{Fig:MR},
the mass--radius relation is shown. The solid lines correspond to the Gibbs EOS and the
dashed lines to the Maxwell construction.
As discussed above, the Gibbs construction softens the EOS at low densities, making
low-mass stars more compact. The solid lines move to smaller values of $R$, and the
desired Z-like relation is more pronounced.
Although the Gibbs construction significantly alters the pressure--density relation at
lower densities, it
does not have a visible effect for stars with higher masses. The maximum mass is almost
the same as in the Maxwell
construction case. For reference, the mass--radius relations of other nuclear models are
also shown. The names of the models are listed in the
legend: APR~\cite{Akmal:1998cf}, DD2, DD2F~\cite{Typel:2014tqa},
BSK20~\cite{Chamel:2011aa},
FSUGarnet \cite{Chen:2014mza}, IUFSU \cite{Fattoyev:2010mx} and
BigApple~\cite{Fattoyev:2020cws}. The latter three are relativistic mean-field models
similar to the one introduced in
this work.

In Fig.~\ref{Fig:MR}, the observational results are also shown.
The 68\% confidence level of the mass--radius measurement for HESS J1731-347 is indicated
by the orange region.
The blue elliptical regions correspond to the most accurate NICER 
measurements of the objects PSR 0740+6620~\cite{Miller:2021qha}, PSR
J0030+0451~\cite{Miller:2019cac}, and PSR J0437-4715~\cite{Choudhury:2024xbk}. The recent
result for PSR J1231-1411~\cite{Qi:2025mpn} is also included, but we mention that it is
debated due to the earlier analysis~\cite{Salmi:2024bss}.
We also include the mass and radius estimates from X-ray bursts from an accreting NS in 4U
1702-429, presented in~\cite{Nattila:2017wtj}.

The interval for the lower bound for the maximum mass, derived from precise Shapiro
delay measurements for PSR J0740+6620, is indicated by the green region.
The gray regions correspond to the two 
estimated masses of the components of the binary system that merged and produced
gravitational waves in the event GW170817.
The black dashed lines mark a lower-bound interval for the \correction{lighter} component 
of the event GW190814 under the assumption that it was a
rapidly spinning NS~\cite{Most:2020bba}. The red dashed regions correspond to estimated
excluded regions from the analysis of
the GW170817 event~\cite{Bauswein:2017vtn,Annala:2017llu}.

All three models, with $\Cs = 12,13,14 \fm2$, fulfill observational constraints, although
the model with $\Cs=14 \fm2$ is in better agreement with the lower bound for the maximum
NS mass based on the pulsar
J0740+6620. The only problematic object is PSR J1231-1411. This is as compact as HESS
J1731-347 and has a mass even larger.
As discussed above, such a small radius is sensitive to the assumptions concerning the
hotspot shape and NS atmosphere modelling~\cite{Qi:2025mpn,Salmi:2024bss}.

\begin{figure}[t]
\includegraphics[width=1\columnwidth]{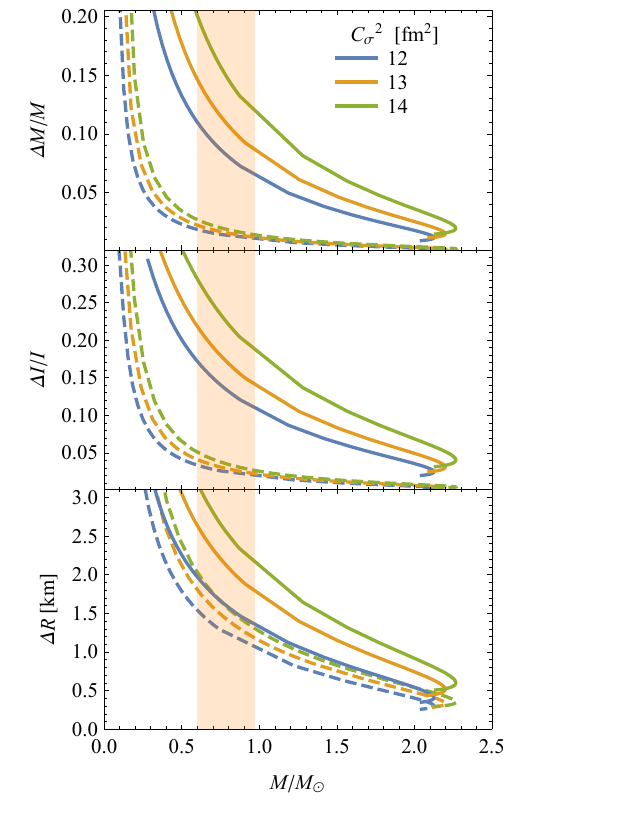}
	\caption{\label{Fig:crust}
Crustal fractions of mass, moment of inertia, and radius. Dashed lines
represent results obtained with the Maxwell construction. The orange shaded region
indicates the mass measurement for HESS J1731‑347.}
\end{figure}

Finally, we describe the main crustal properties in the CCT model. 
Applying the Gibbs conditions to the two-phase system that appears in our model reveals
that the NS crust may
encompass a much wider range of densities than in most models. If $n_{2ph}$ is treated as
the crust--core transition
density, {$n_{cc}$}, the NS crust reaches up to approximately $2n_0$ (see
Table~\ref{nc-tab}). This means that the crustal contribution to the
stellar parameters is much more pronounced. In Fig.~\ref{Fig:crust}, the crust thickness
$\Delta R$ and its contribution to the stellar
mass and moment of inertia are shown. The dashed line indicates the crustal parameters
derived from the CCT model with the Maxwell construction, which was used
in~\cite{Kubis:2023gxa}, where the crust--core transition was at around $n_{cc}=0.056
\fm2$. Their
behavior does not diverge substantially from most of the other models, but the values
obtained with the Gibbs construction are considerably different.
It is interesting that for the case of HESS J1731-347, the crust is about 50\% thicker.
Much greater differences are obtained for crustal mass and moment of inertia. 
The crustal contribution $\Delta M/M$ is between 7\% and 20\%, depending on the total mass
and stiffness of the EOS, and
similarly contribution to the total moment of inertia $\Delta I/I$ takes values from 12\%
to 27\%.
The large values of $\Delta I/I$ are specially interesting in the context of pulsar
glitching observations.
Estimation of crustal moment of inertia for the strongest glitcher, Vela pulsar, 
suggested that $\Delta I/I > 1.4\%$ \correction{which} is easily
satisfied by most of EOSs. However, Andersson {\it et.al} \cite{Andersson:2012iu} have
shown that inclusion of
crustal entrainment requires much thicker crust, i.e. $\Delta I/I > 6 - 10\%$.   
The bound is easily satisfied in our model. For low mass star the crustal moment of
inertia reaches 25\%. Of course it declines with
stellar mass: for canonical mass $1.4 M_\odot$ it is 7-15\%, but even for the most massive
ones it is still around 5-10\%.
The thick and massive crust in neutron star is not common, however was already suggested
in the works \cite{Piekarewicz:2014lba}
and \cite{Fortin:2016hny}. In the work \cite{Piekarewicz:2014lba} authors remarked that
the high crust-core transition pressure
$P_t \approx 0.3\!-\!0.7 ~\rm MeV/fm{^3}$ is crucial in supporting the thick crust. In our
model the transition pressure
(corresponding to the point where the two-phase system vanishes) are even higher: $P_t =
1.50, 1.79, 2.17 ~\rm MeV/fm{^3}$ for the
three  different $\Cs$.

\section{Conclusions}

We have shown that using the thermodynamically consistent Gibbs construction for the
first-order phase transition that
occurs in the relativistic mean-field model with $\sigma$--$\delta$ and $\omega$--$\delta$
meson crossing term leads to a thick
neutron \correction{star} crust. The mass--radius relation may be controlled to some extent by the coupling
to the $\sigma$ field,
and in a reasonable range of its values, the mass--radius relation is in close agreement
with the most reliable observational
constraints.
It is worth emphasizing the thick crust together with large compactness of the star leads
to a large contribution to the
crustal moment of inertia and significantly alters analysis of rotational parameters. 

The thermal relaxation time depends like $\Delta R^2$ \cite{Yakovlev:2000jp}, so the crust
thickness
is also relevant for thermal relaxation after outbursts in low-mass X-ray binaries
\cite{Chaikin:2018zbc} and neutron star cooling.
The solid conclusion requires, however, inclusion of various neutrino emission types and
finite size effects for cluster in two-phase
system. Similarly, other phenomena like glitch activity, post-quake behaviour, crust
oscillation and GW emission from deformed
crust should be affected by its large thickness. The description of internal structure and
elastic properties of the crust
demands going beyond the simple two-phase system analysis done here. Its extension is in
progress and will allow for correct
predictions.

\end{document}